\documentclass[11pt]{article}

\usepackage{a4wide}
\usepackage{wrapfig}
\usepackage{floatflt}
\usepackage{amssymb}
\usepackage{amsmath}
\usepackage{graphicx}
\usepackage{xspace}
\usepackage{url}

\newcommand{\comment}[1]{\textbf{[#1]}}

\newcommand{\as}{\alpha_s}

\newcommand{\order}[1]{{\mathcal O}\left(#1\right)}

\newcommand{\eg}{e.g.\ }

\newcommand{\JA}{\text{JA}}
\newcommand{\akt}{\text{anti-}k_t}

\newcommand{\GeV}{\,\text{GeV}}
\newcommand{\TeV}{\,\text{TeV}}

\title{
  \textbf{The anti-$\boldsymbol{k_t}$ jet clustering algorithm}
}

\author{
  Matteo~Cacciari and Gavin~P.~Salam\\
  \small \it LPTHE\\
  \it\small 
  UPMC Universit\'e  Paris 6,\\
  \it\small 
  Universit\'e Paris Diderot -- Paris 7,\\ 
  \it\small 
  CNRS UMR 7589, Paris, France\\[10pt]
  Gregory Soyez\\
  \small \it  Brookhaven National Laboratory, Upton, NY, USA
}

\date{}

\begin{document}


\maketitle

\vspace{-8.5cm}
\begin{flushright}
  February 2008\\
  LPTHE-07-03\\
\end{flushright}
\vspace{6.3cm}
 
\noindent \textbf{Abstract: }{%
  The $k_t$ and Cambridge/Aachen inclusive jet finding algorithms for
  hadron-hadron collisions can be seen as belonging to a broader class
  of sequential recombination jet algorithms, parametrised by the
  power of the energy scale in the distance measure. We examine some
  properties of a new member of this class, for which the power is
  negative. This ``anti-$k_t$'' algorithm essentially behaves like an
  idealised cone algorithm, in that jets with only soft fragmentation
  are conical, active and passive areas are equal, the area anomalous
  dimensions are zero, the non-global logarithms are those of a rigid
  boundary and the Milan factor is universal. 
  None of these properties hold for existing sequential recombination
  algorithms, nor for cone algorithms with split--merge steps, such as
  SISCone. They are however the identifying characteristics of the
  collinear unsafe plain ``iterative cone'' algorithm, for which the
  anti-$k_t$ algorithm provides a natural, fast, infrared and
  collinear safe replacement.
} \bigskip \bigskip
\bigskip

\section{Introduction and definition}

Jet clustering algorithms are among the main tools for analysing data
from hadronic collisions. Their widespread use at the Tevatron and the
prospect of unprecedented final-state complexity at the upcoming 
Large Hadron Collider (LHC) 
have stimulated considerable debate concerning the merits of different
kinds of jet algorithm.
Part of the discussion has centred on the relative advantages of
sequential recombination ($k_t$ \cite{kt} and
Cambridge/Aachen~\cite{cam}) and cone (\eg \cite{Blazey})
jet algorithms, with
an issue of particular interest being that of the regularity of the
boundaries of the
resulting jets. This is related to the question of their sensitivity
to non-perturbative effects like hadronisation and underlying event
contamination and arises also in the context of experimental
calibration. 

Recently \cite{css-area}, tools have been developed that allow one, for the
first time, to support the debate with analytical calculations of the
contrasting properties of boundaries of jets within different algorithms.
One of the main results of that work is that
all known infrared and
collinear (IRC) safe algorithms have the property that soft radiation
can provoke irregularities in the boundaries of the final jets. This is the case even
for SISCone~\cite{siscone}, an IRC-safe jet algorithm based on the
search for stable cones, together with a split--merge step that
disentangles overlapping stable cones. One might describe current
IRC-safe
algorithms in general as having a `soft-adaptable' boundary.

A priori it is not clear whether it is better to have regular
(`soft-resilient') or less regular (soft-adaptable) jets. In particular,
regularity implies a certain rigidity in the jet algorithm's
ability to adapt a jet to the successive branching nature of QCD
radiation. On the other hand knowledge of the typical shape of jets is
often quoted as facilitating experimental calibration of jets, and
soft-resilience can simplify certain theoretical calculations, as well
as eliminate some parts of the momentum-resolution loss caused by
underlying-event and pileup contamination.

Examples of jet algorithms with a soft-resilient boundary are the
plain ``iterative cone'' algorithm, as used for example in the CMS
collaboration \cite{CMSTDR}, and fixed-cone algorithms such as Pythia's
\cite{Pythia} CellJet. The CMS iterative cone  
takes the hardest object (particle, calorimeter tower) in the event,
uses it to seed an iterative process of looking for a stable cone,
which is then called a jet. It then removes all the particles
contained in that jet from the event and repeats the procedure with
the hardest available remaining seed, again and again until no seeds
remain. 
The fixed-cone algorithms are similar, but simply define a jet as the
cone around the hardest seed, skipping the iterative search for a
stable cone.
Though simple experimentally, both kinds of algorithm have the crucial drawback that if
applied at particle level they are collinear unsafe, since the hardest
particle is easily changed by a quasi-collinear splitting, leading to
divergences in higher-order perturbative calculations.\footnote{%
  This is discussed in the appendix in detail for the iterative cone,
  and there we also introduce
  the terminology iterative cone with split--merge steps (IC-SM) and
  iterative cone with progressive removal (IC-PR), so as to distinguish the two
  broad classes of iterative cone algorithms.}

In this paper it is not our intention to advocate one or other type of
algorithm in the debate concerning soft-resilient versus
soft-adaptable algorithms.  Rather, we feel that this debate can be
more fruitfully served by proposing a simple, IRC
safe, soft-resilient jet algorithm, one that leads to jets whose shape
is not influenced by soft radiation.
To do so, we take a quite non-obvious route, because instead of
making use of the concept of a stable cone, we start by
generalising the existing sequential recombination
algorithms, $k_t$~\cite{kt} and Cambridge/Aachen~\cite{cam}.

As usual, one 
introduces 
distances $d_{ij}$ between entities (particles, pseudojets)
$i$ and $j$ and $d_{iB}$ between entity $i$ and the beam (B). The
(inclusive) clustering proceeds by identifying the smallest of the
distances and if it is a $d_{ij}$ recombining entities $i$ and $j$,
while if it is $d_{iB}$ calling $i$ a jet and removing it from the list
of entities.  The distances are recalculated and the procedure
repeated until no entities are left.

The extension relative to the $k_t$ and Cambridge/Aachen algorithms
lies in our definition of the distance measures:
\begin{subequations}
  \label{eq:genkt}
  \begin{align}
    d_{ij} &= \min(k_{ti}^{2p}, k_{tj}^{2p}) \frac{\Delta_{ij}^2}{R^2}\,,\\
    d_{iB} &= k_{ti}^{2p}\,,
  \end{align}
\end{subequations}
where $\Delta_{ij}^2 = (y_i-y_j)^2 + (\phi_i - \phi_j)^2$ and
$k_{ti}$, $y_i$ and $\phi_i$ are respectively the transverse momentum,
rapidity and azimuth of particle $i$. 
In addition to the usual radius parameter $R$, we have added a
parameter $p$ to govern the relative power of the energy versus
geometrical ($\Delta_{ij}$) scales.

For $p=1$ one recovers the inclusive $k_t$ algorithm. It can be shown
in general  that for $p>0$ the behaviour of the jet algorithm with respect
to soft radiation is rather similar to that observed for the $k_t$
algorithm, because what matters is the ordering between particles and
for finite $\Delta$ this is maintained for all positive values of
$p$.
The case of $p=0$ is special and it corresponds to the inclusive
Cambridge/Aachen algorithm.

Negative values of $p$ might at first sight seem pathological. We
shall see that they are not.\footnote{Note that, for $p<0$,
  $\min(k_{ti}^{2p}, k_{tj}^{2p})$ differs from another possible
  extension, $[\min(k_{ti}^{2}, k_{tj}^{2})]^p$, which \emph{can} lead
  to strange behaviours.%
} %
The behaviour with respect to soft radiation will be
similar for all $p<0$, so here we will concentrate on $p=-1$, and
refer to it as the ``anti-$k_t$'' jet-clustering algorithm.

\section{Characteristics and properties}

\subsection{General behaviour}
The functionality of the anti-$k_t$
algorithm can be understood by considering an event
with a few well-separated hard particles with transverse momenta
$k_{t1}$, $k_{t2}$, \ldots and many soft particles. The $d_{1i} =
\min(1/k_{t1}^{2}, 1/k_{ti}^{2}) \Delta_{1i}^2/R^2$ between a hard
particle $1$ and a soft particle $i$ is exclusively determined by the
transverse momentum of the hard particle and the $\Delta_{1i}$
separation. The $d_{ij}$ between similarly separated soft particles
will instead be much larger. Therefore soft particles will tend to
cluster with hard ones long before they cluster among themselves. If a
hard particle has no hard neighbours within a distance $2R$, then it
will simply accumulate all the soft particles within a circle of radius
$R$, resulting in a perfectly conical jet.

If another hard particle $2$ is present such that $R < \Delta_{12} <
2R$ then there will be two hard jets. It is not possible for both to
be perfectly conical. If $k_{t1} \gg k_{t2}$ then jet $1$ will be
conical and jet $2$ will be partly conical, since it will miss the
part overlapping with jet $1$. Instead if $k_{t1} = k_{t2}$ neither
jet will be conical and the overlapping part will simply be divided by
a straight line equally between the two. 
For a general situation, $k_{t1} \sim k_{t2}$, both cones will be
clipped, with the boundary $b$ between them defined by $\Delta
R_{1b}/k_{t1} = \Delta_{2b}/k_{t2}$.

Similarly one can work out what happens with $\Delta_{12} < R$. Here
particles $1$ and $2$ will cluster to form a single jet. If $k_{t1}
\gg k_{t2}$ then it will be a conical jet centred on $k_1$. For
$k_{t1} \sim k_{t2}$ the shape will instead be more complex, being the
union of cones (radius $< R$) around each hard particle plus a cone
(of radius $R$) centred on the final jet.

The key feature above is that the soft particles do not modify the
shape of the jet, while hard particles do. I.e.\ the jet boundary in
this algorithm is resilient with respect to soft radiation, but flexible
with respect to hard radiation.\footnote{For comparison, IC-PR algorithms
  behave as follows: with $R < \Delta_{12} < 2R$, the harder of the
  two jets will be fully conical, while the softer will be clipped
  regardless of whether $p_{t1}$ and $p_{t2}$ are similar or disparate
  scales; with $\Delta_{12} < R$ the jet will be just a circle
  centred on the final jet.}

\begin{figure}[t]
  \centering
  \includegraphics[width=0.49\textwidth]{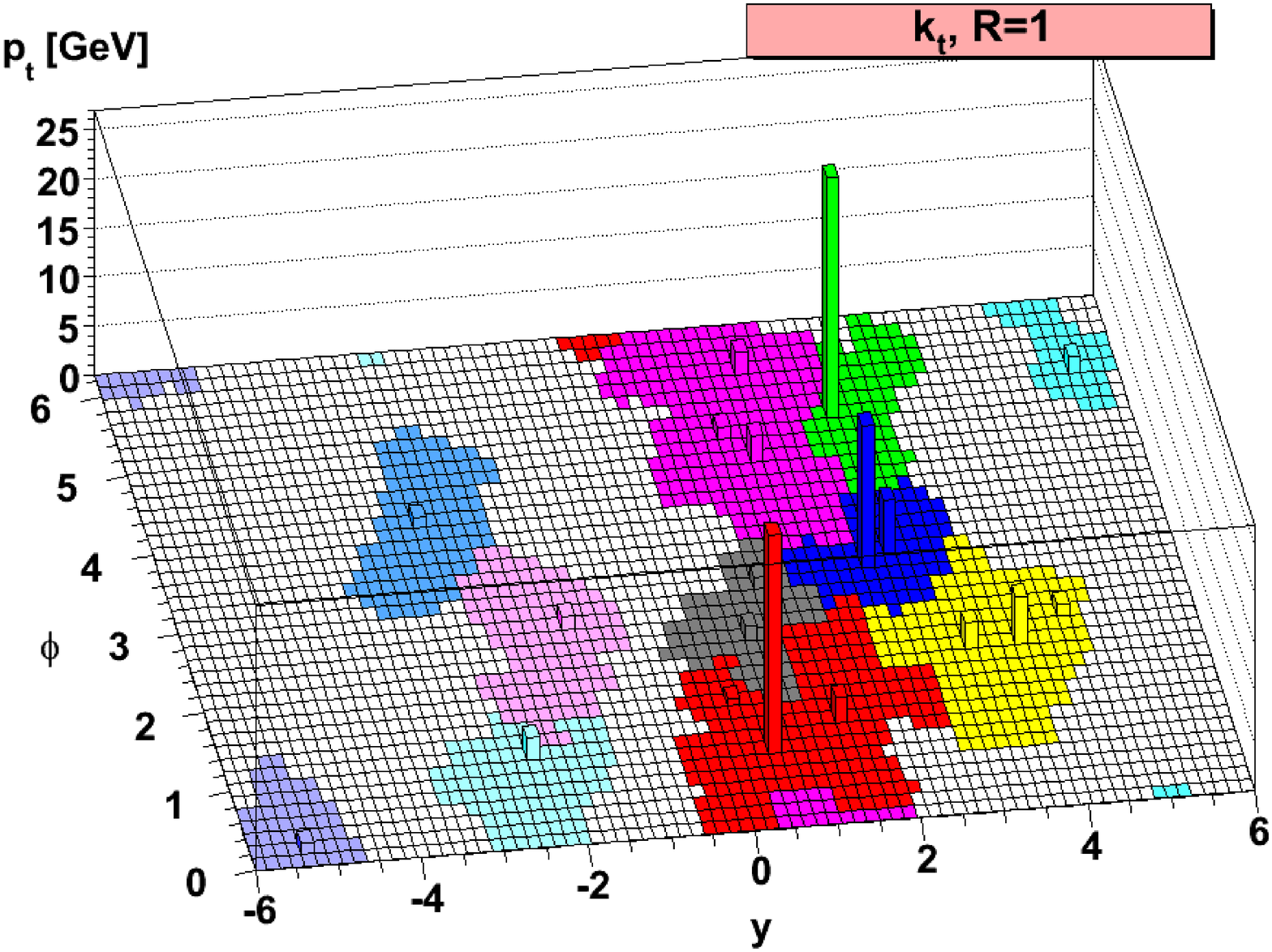}%
  \hfill
  \includegraphics[width=0.49\textwidth]{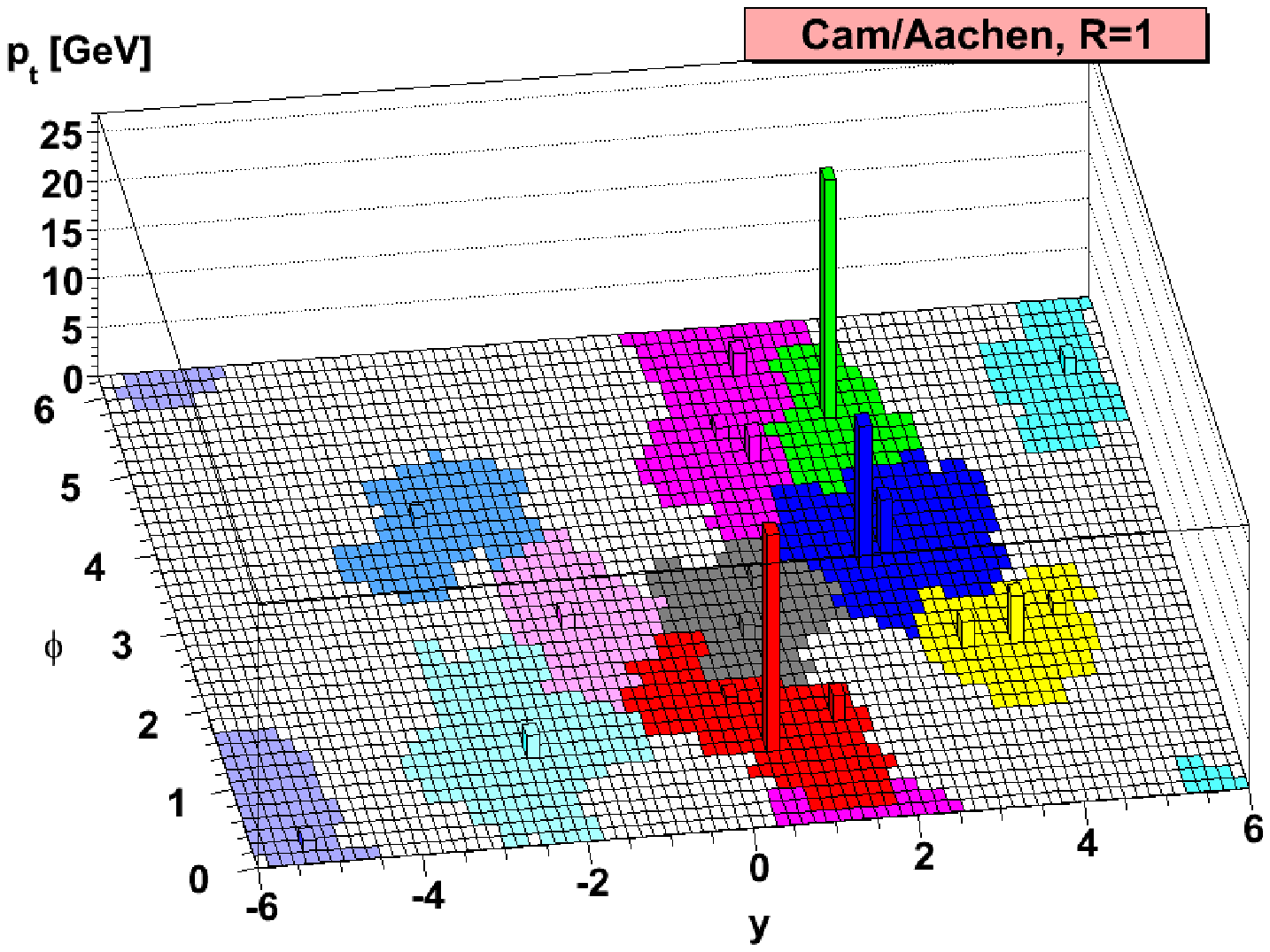}%
  \\
  \includegraphics[width=0.49\textwidth]{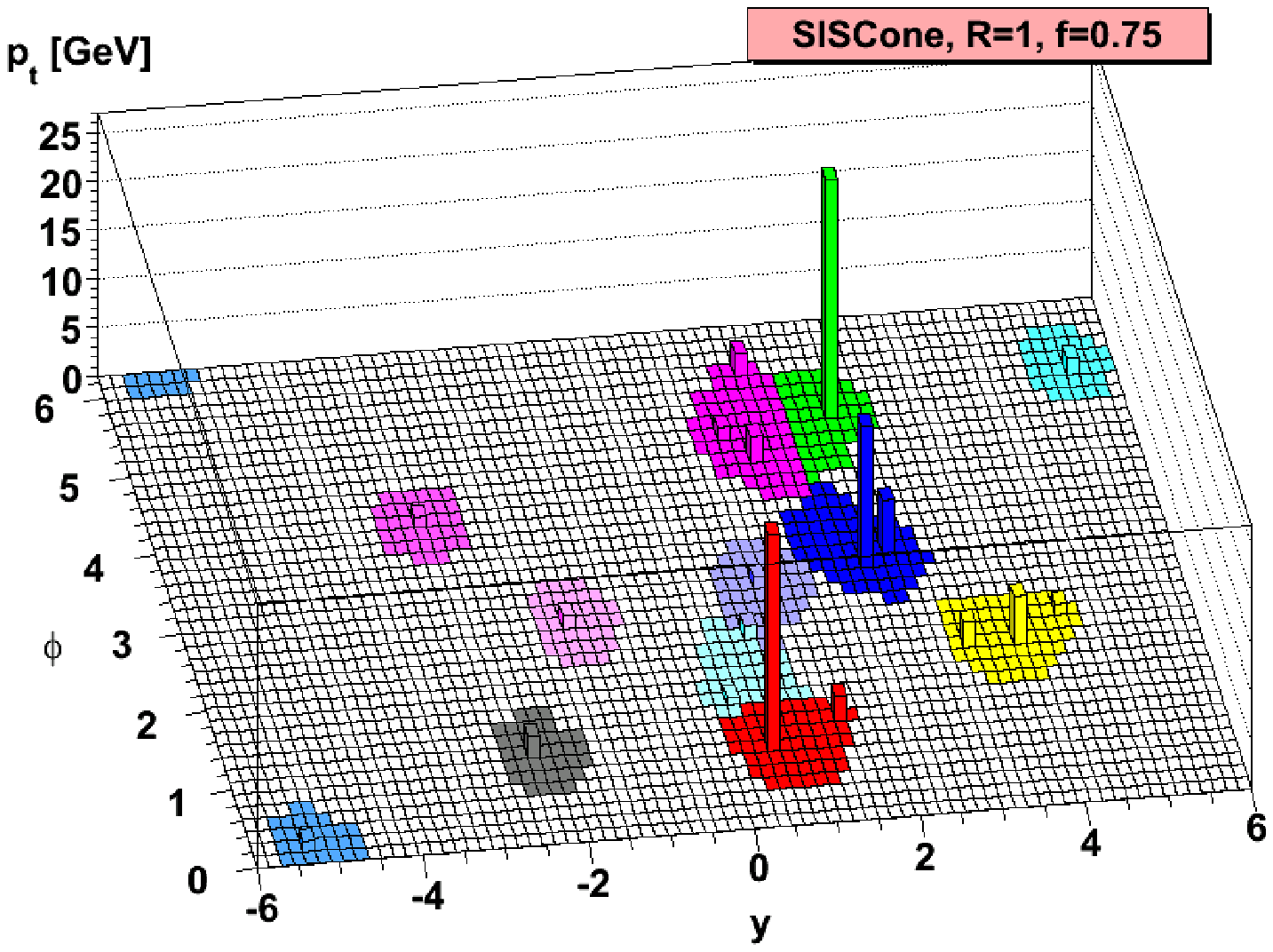}%
  \hfill
  \includegraphics[width=0.49\textwidth]{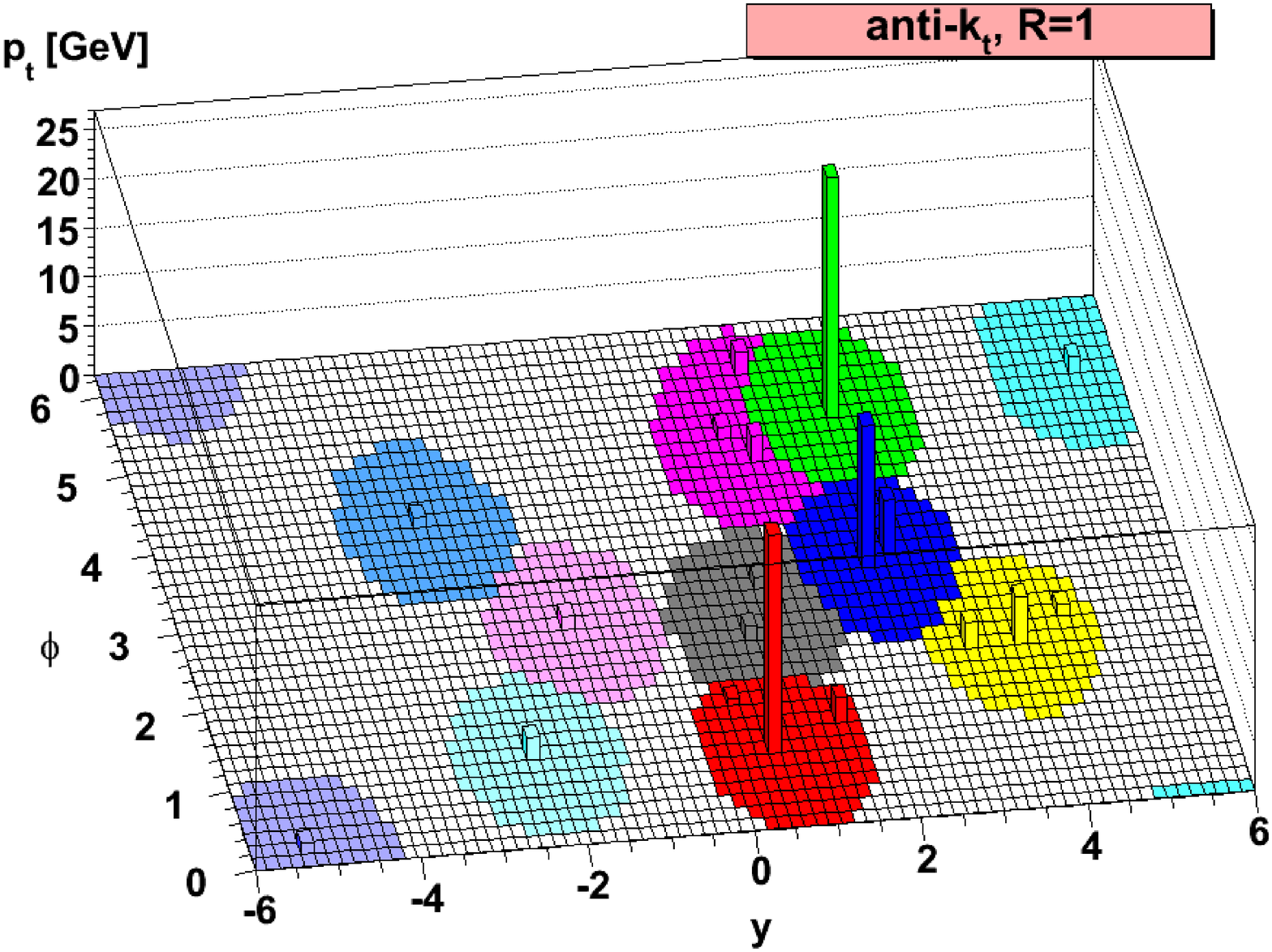}%
  \caption{A sample parton-level event (generated with
    Herwig~\cite{Herwig}), together with many random soft ``ghosts'',
    clustered with four different jets algorithms, illustrating the
    ``active'' catchment areas of the resulting hard jets. For $k_t$
    and Cam/Aachen the detailed shapes are in part determined by the
    specific set of ghosts used, and change when the ghosts are
    modified.}
  \label{fig:4algs}
\end{figure}

The behaviours of different jet algorithms are illustrated in
fig.~\ref{fig:4algs}. We have taken a parton-level event together with
$\sim 10^4$ random soft `ghost' particles (as in \cite{css-area}) and
then clustered them with 
4 different jet algorithms. For each of the partonic jets, we have
shown the region within which the random ghosts are clustered into
that jet. For the $k_t$ and Cambridge/Aachen algorithms, that region
depends somewhat on the specific set of ghosts and the jagged borders
of the jets are a consequence of the randomness of the ghosts --- the
jet algorithm is adaptive in its response to soft particles, and that
adaptiveness applies also to the ghosts which take part in the
clustering. For SISCone one sees that single-particle jets are regular
(though with a radius $R/2$ --- as pointed out in~\cite{css-area}),
while composite jets have more varied shapes.  Finally with the
anti-$k_t$ algorithm, the hard jets are all circular with a radius
$R$, and only the softer jets have more complex shapes. The pair of
jets near $\phi=5$ and $y=2$ provides an interesting example in this
respect. The left-hand one is much softer than the right-hand
one. SISCone (and Cam/Aachen) place the boundary between the jets
roughly midway between them. Anti-$k_t$ instead generates a circular
hard jet, which clips a lens-shaped region out of the soft one,
leaving behind a crescent. 

The above properties of the anti-$k_t$ algorithm translate into
concrete results for various quantitative properties of jets, as we
outline below.

\subsection{Area-related properties} 
The most concrete context in which to
quantitatively discuss the properties of jet boundaries for different
algorithms is in the calculation of jet areas.

Two definitions were given for jet areas in
\cite{css-area}: the passive area ($a$) which measures a jet's susceptibility
to point-like radiation, and the active area ($A$) which measures its
susceptibility to diffuse radiation.
The simplest place to observe the impact of soft resilience is in the
passive area for a jet consisting of a hard particle $p_1$ and a
soft one $p_2$, separated by a $y-\phi$ distance $\Delta_{12}$. In
usual IRC safe jet algorithms (JA), the passive area 
$a_{\JA,R}(\Delta_{12})$ is $\pi R^2$ when
$\Delta_{12} = 0$, but changes when
$\Delta_{12}$ is increased. In contrast, since the boundaries of
anti-$k_t$ jets are unaffected by soft radiation, their passive area
is always independent of $\Delta_{12}$:
\begin{equation}
  \label{eq:passive}
  a_{\text{anti-}k_t,R}(\Delta_{12}) = \pi R^2\,.
\end{equation}
Furthermore, since soft particles only cluster among themselves
\emph{after} all clusterings with hard particles have occurred,
passive and active areas are identical, another feature that is unique
to the anti-$k_t$ algorithm.
Thus, specifically for our energy-ordered two-particle configuration,
the active area is
\begin{equation}
  \label{eq:active}
  A_{\text{anti-}k_t,R}(\Delta_{12}) =
  a_{\text{anti-}k_t,R}(\Delta_{12})  = \pi R^2\,.
\end{equation}
In~\cite{css-area} the fact that $a_{\JA,R}(0) \ne
a_{\JA,R}(\Delta_{12})$ (and similarly for the active area) meant that
the jet areas acquired an anomalous dimension in their dependence on the
jet transverse momentum $p_{tJ}$, related to emission of perturbative
soft particles:
\begin{equation}
  \label{eq:anom-dim}
  \langle a_{\JA,R} \rangle = \pi R^2 +  d_{\text{JA},R}\,\frac{C_1}{\pi b_0} \ln
  \frac{\alpha_s(Q_0)}{\alpha_s(R p_{tJ})} \,,
\end{equation}
where $C_1$ is the colour factor of the hard particle in the jet,
$Q_0$ is the non-perturbative scale and the coefficient
of the anomalous dimension, $d_{\JA,R}$, is given by
\begin{equation}
  \label{eq:d}
  d_{\text{JA},R} = \int_0^{2R} \frac{d\theta}{\theta} (
  a_{\text{JA},R}(\theta) - \pi R^2)\,.
\end{equation}
Obviously in this case $d_{\akt,R}=0$ and similarly for the active
area anomalous dimension coefficient $D_{\akt,R}$.
One corollary of this is that $\akt$ jet areas can be calculated
perturbatively, order by order, since they are infrared safe. In this
respect we recall that they do deviate from $\pi R^2$ for
configurations with several neighbouring hard particles.

\begin{table}\small
  \newcommand{\D}[1]{\comment{$#1$}}
  \centering   
  \begin{tabular}{r|cc|cc|cc|cc}
                 & $a$(1PJ) & $A$(1PJ)  & $\sigma$(1PJ) & $\Sigma$(1PJ)  & $d$         &  $D$   & $s$     & $S$       \\\hline 
   $k_t$         & $1$      & $0.81$    & $0$           & $0.28$         & $\,\,0.56$  & $0.52$ & $0.45$  & $0.41$  \\ \hline  
   Cam/Aachen    & $1$      & $0.81$    & $0$           & $0.26$         & $\,\,0.08$  & $0.08$ & $0.24$  & $0.19$  \\ \hline  
   SISCone       & $1$      & $1/4$    & $0$           & $0$            & $\!\!-0.06$ & $0.12$ & $0.09$  &  $0.07$   \\ \hline 
   anti-$k_t$    & $1$      & $1$       & $0$           & $0$            & $0$         & $0$    & $0$     &  $0$      \\ \hline
  \end{tabular}
  \caption{A summary of main area results for the anti-$k_t$ algorithm
    compared to those for other IRC safe algorithms in
    \cite{css-area}: the passive ($a$) and active ($A$) areas for
    1-particle jets (1PJ), the magnitude of the passive/active area
    fluctuations ($\sigma$, $\Sigma$), followed by the coefficients of
    the respective anomalous dimensions ($d$, $D$; $s$, $S$), in the
    presence of perturbative QCD radiation. All results are normalised
    to $\pi R^2$, and rounded to two decimal figures.  
    For algorithms other than anti-$k_t$, active-area
    results hold only in the small-$R$ limit, though finite-$R$
    corrections are small.}
  \label{tab:summary}
\end{table}

\begin{figure}
\centerline{\includegraphics[angle=270,width=\textwidth]{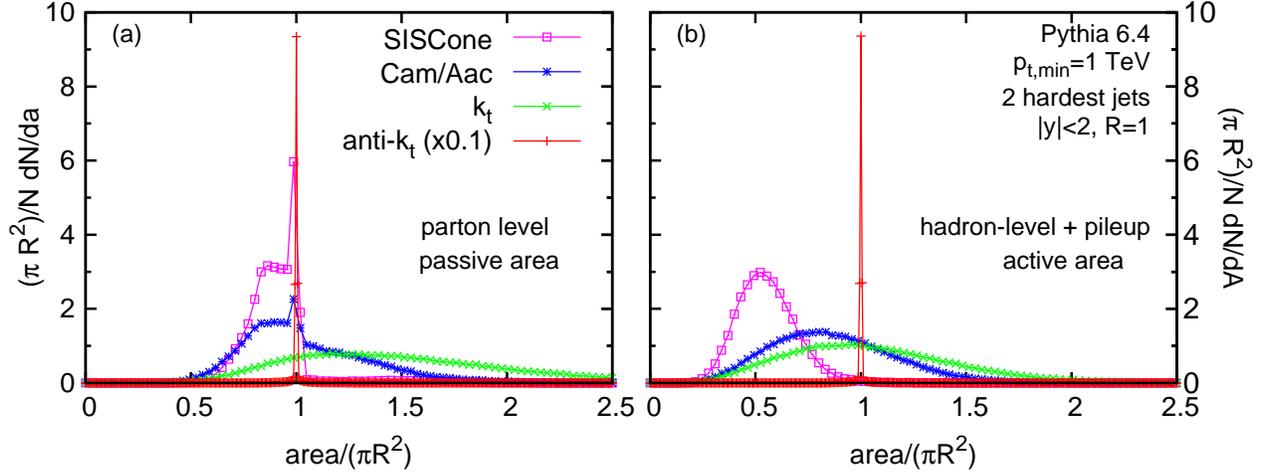}}
\caption{Distribution of areas in dijet events at the LHC. We have generated
  events with Pythia 6.4 with a $p_{t,min}$ of 1 TeV. Only the two
  hardest jets have been kept with a further requirement $|y|\le 2$. The area distribution
  obtained from
  anti-$k_t$ (scaled by 0.1) is compared to the other algorithms. (a) passive area at
  parton level; (b) active area at hadron level including the
  underlying event and pileup corresponding to high
  luminosity LHC running.}
\label{fig:histogram}
\end{figure}

\begin{figure}
\centerline{\includegraphics[width=0.5\textwidth]{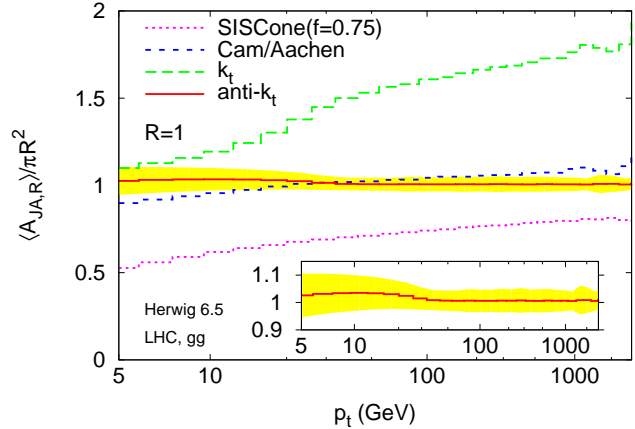}}
\caption{Average jet area in dijet events at the LHC. We have generated events
  with Herwig 6.5 and only the two hardest jets with $|y|\le 2$ have
  been kept. The curves correspond to the average jet area at a given
  $p_t$. The yellow band around the anti-$k_t$ line corresponds to the
  area fluctuations. For clarity, the latter are not shown for the
  other algorithms. The encapsulated graph is a zoom on the anti-$k_t$
  results.
  Note that the horizontal scale is $\ln \ln(p_t/\Lambda_{QCD})$ with
  $\Lambda_{QCD}=200$~MeV.}
\label{fig:anomdim}
\end{figure}

A summary of these and other properties of the jet area is given in
table~\ref{tab:summary}, together with a comparison to other IRC safe
algorithms as studied in~\cite{css-area}.  
Fig.~\ref{fig:histogram} illustrates the distribution of
areas in dijet events at the LHC (generated with Pythia 6.4, with a transverse
momentum cut of $1 \TeV$ on the $2\to 2$ scattering, considering those
among the two hardest jets in each event that have $|y| < 2$),
compared with those for
other algorithms, and one sees a near $\delta$-function at an area of
$\pi R^2$, to be compared to the broader distributions of other
algorithms. Fig.~\ref{fig:anomdim} shows the mean jet area, together
with a band representing the event-by-event fluctuations of the area,
as a function of the jet $p_t$ in $gg \to gg$ events, now generated with
Herwig. This
highlights the independence of the area on the jet $p_t$ and, once
again, the very small fluctuations in the jet area.
In this respect, we recall that the impact of the UE and pileup on the
momentum resolution for jets is related both to the typical value of
the area (smaller than $k_t$, similar to Cambridge/Aachen, larger than
SISCone) and to the extent of the fluctuations of the area, which are
close to zero only for anti-$k_t$.

\paragraph{Back reaction.} Suppose one has a hard scattering event
that leads to a set of jets $\{J_i\}$. If one adds a soft event (UE,
pileup) to it and reruns the the algorithm, one will obtain a set of
jets $\{\tilde J_i\}$ that differ in two respects: 
soft energy will have been added to each jet, and additionally the way
the particles from the hard event are distributed into jets may also
have changed: even if one finds the $\tilde J_i$ that is closest to
the original $J_1$, the two jets will not contain exactly the same
subset of particles from the original hard event. This was called
``back-reaction'' in \cite{css-area}.
If the soft particles that are added have a density $\rho$ of transverse
momentum per unit area, then for usual sequential
recombination algorithms the probability, $dP/d\ln \Delta p_t^{(B)}$, of
having a back reaction of $\Delta p_t^{(B)}$ is $\order{\smash{\as
    \rho/\Delta p_t^{(B)}}}$ for $\Delta p_t^{(B)} \gtrsim \rho$.\footnote{For smaller
values of $\Delta p_t^{(B)}$ the full answer requires a resummation that has
yet to be carried out, and in the case of SISCone, it depends on the
nature of the fluctuations of soft event.}

\begin{figure}
  \centering
  \includegraphics[width=0.5\textwidth]{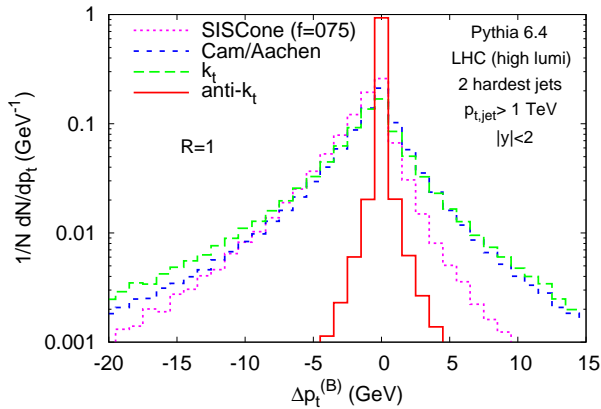}
  \caption{Distribution of back-reaction for the anti-$k_t$ algorithm
    as compared to $k_t$, Cambridge/Aachen and SISCone. It is
    calculated for dijet events simulated with Pythia 6.4 in which the
    two hardest jets have $p_t > 200 \GeV$ and are both situated at
    $|y|<2$. The back reaction corresponds to the net transverse
    momentum change of each of the two hardest jets due to the
    reassignment of non-pileup particles when one adds high-luminosity
    LHC pileup to the event ($\sim$ 25 $pp$ interactions per bunch
    crossing). }
  \label{fig:back-reac}
\end{figure}

For the anti-$k_t$ algorithm, one can show that the probability of
back-reaction is suppressed not by the amount of back reaction itself, $\Delta
p_t^{(B)}$, but by the jet transverse momentum $p_{tJ}$, leading to
a much smaller effect. The impact of this is illustrated in
fig.~\ref{fig:back-reac}, which shows the back reaction that occurs
for the hard jet events when adding high luminosity LHC pileup to
the event. One sees that it is strongly suppressed for the anti-$k_t$
algorithm relative to the others, a characteristic that can help reduce
the smearing of jets' momenta due to the UE and pileup.

\subsection{Other properties}

\paragraph{Non-global logarithms.} Resummations of observables
involving boundaries in phase space are known to involve `non-global'
logarithms. Examples are the jet-mass
distribution~\cite{DasSalJetMass} and energy flow distributions in
restricted regions~\cite{DasSalEFlow,BKS,BMS}. In both cases there are
all-order single logarithmic terms (for a hard scale $Q$, $\as^n \ln^n
Q/m$ for jet-mass distributions, $\as^n \ln^n Q/E$ for energy flow),
that are related to the impact of emissions outside the boundary that
radiate a gluon across the boundary and so affect the observable
defined only on particles within the boundary.

It was pointed out in \cite{ApplebySeymour} that the structure of
these non-global terms is more complex with clustering jet algorithms,
because the boundary of the jet itself depends on the soft
radiation.\footnote{There are cases where, despite the added
  analytical complexity, this is an advantage since it can cancel a
  significant part of the non-global logarithms.}  Furthermore this
affects even the contribution from independent
emissions~\cite{BanfiDasgupta} which remained simple in the case of a
rigid boundary. So far these effects have been calculated only for the
$k_t$ algorithm, while for the cone and Cambridge/Aachen algorithms
little is known about the non-global logarithms other than that they
too involve a subtle interaction between the clustering and the
non-global resummation.

Because soft radiation does not affect the boundary of $\akt$ jets, it
is straightforward to see that their single-logarithmic non-global
terms are identically those associated with ideal cones, considerably
simplifying their determination.

\paragraph{Milan factor.} A crucial ingredient in analytical studies of
non-perturbative effects in event and jet-shapes is the Milan
factor~\cite{DLMS,DasMagSmy,BenBroMag}, which is the correction that
relates calculations made in a single soft (almost non-perturbative)
gluon approximation to calculations in which the soft gluon splits at
large angle. A remarkable characteristic of the Milan factor
(`universality') is that it is identical for all event shapes. This is
essentially because for all event shapes, if one takes an event with
hard momenta $p_i$ and soft momenta $k_i$, then the event shape can be
approximated as
\begin{equation}
  \label{eq:MilanApprox}
  V(\{p_i\},\{k_i\}) = V(\{p_i\}) + \sum_{\{k_i\}}
  f_V(\theta_i,\{p_i\}) k_{ti} 
\end{equation}
where $f_V$ is a function specific to the event shape observable $V$.
The key feature is the linearity of the second term on the
right-hand side of eq.~(\ref{eq:MilanApprox}) (see also
\cite{LeeSterman}). If, however, the event
shape is defined for just a jet (or is simply a characteristic of the
jet such as its transverse momentum), then one loses the linear
dependence on soft momenta, since the question of whether one soft
particle contributes to the observable is affected by its potential
clustering with other soft particles.

In the case of $\akt$ jets, the independence of the jet boundary on the
soft particle configuration means that the approximation
eq.~(\ref{eq:MilanApprox}) does hold and the Milan factor retains its
`universal' value 
($M=1.49$ for $3$ active non-perturbative flavours~\cite{DLMS,DasMagSmy}).

\paragraph{Speed.} A relevant issue in order for a jet algorithm to be
useful in practice is the computing time required to carry out the
clustering. The full class of generalised $k_t$ algorithms is amenable
to fast implementation using the techniques of \cite{fastjet}, with
the proviso that for $p<0$, the specific manner in which particles are
clustered triggers a worst-case scenario for the Voronoi-diagram based
dynamic nearest-neighbour graph determination \cite{CGALTriang}. This
means that asymptotically the algorithm takes a time $\order{N^{3/2}}$
to cluster $N$ particles (rather than $N\ln N$ for the $k_t$
algorithm). However for $N\lesssim 20000$ the FastJet
implementation~\cite{FastJetWeb} in any case uses other strategies,
which are insensitive to this issue, and the anti-$k_t$ clustering is
then as fast as $k_t$ clustering.

\subsection{Example application: top reconstruction}
\label{sec:top}

\begin{figure}[t]
  \centering
  \includegraphics[width=0.5\textwidth]{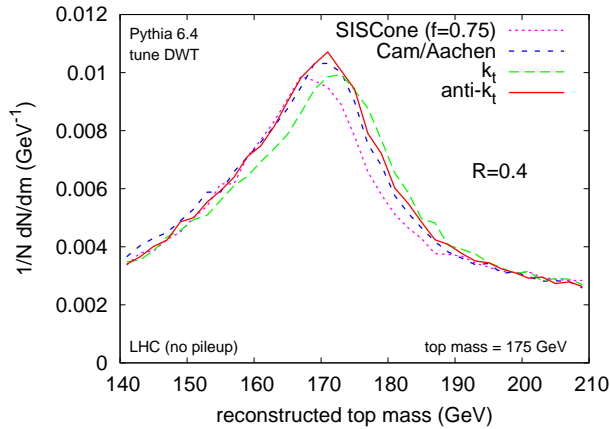}
  \caption{Top mass reconstruction in Pythia-simulated LHC $t\bar t$
    events. Both the $t$ and the $\bar t$ decay hadronically,
    $t \to bW^+ \to bq\bar q$ and $\bar t \to \bar bW^- \to \bar
    bq\bar q$. All jet algorithms have been used with $R=0.4$.}
  \label{fig:top}
\end{figure}

One may wonder whether the unusual soft-resilience of the anti-$k_t$
algorithm leads to poorer results in phenomenological applications.
We have investigated various examples and found that in general this
is not the case. In figure~\ref{fig:top} we illustrate this for top
mass reconstruction in LHC $t\bar t$ events, as simulated with
Pythia~\cite{Pythia}, where both the $t$ and the $\bar t$ decay
hadronically, according to $t \to bW^+ \to bq\bar q$ and $\bar t
\to \bar bW^- \to \bar bq\bar q$. The following simple analysis
procedure has been used: we select events with at least 6 hard jets (with $p_t$ above 10 GeV and
$|y|<5$); we assume that both $b$-jets have been tagged; the 4
hardest remaining jets are paired according to the combination that better
reproduces the $W$ masses; finally, the $W$- and $b$-jets are
recombined to minimise the mass-difference between the two $t$-jets.
We use the same four algorithms shown in figure~\ref{fig:4algs}, now
with $R=0.4$, and find that they all behave rather similarly, with
Cambridge/Aachen and $\akt$ performing marginally better than the other two.
We note that the difference between various choices of $R$ can be
substantially greater than the differences between the various
algorithms at a given $R$. One should also bear in mind that top
reconstruction, near threshold, with a moderate jet radius and no
pileup is a relatively simple application for inclusive jet
algorithms. One expects that greater differences between algorithms
will be seen in other contexts, an extreme example being multi-scale
processes with large pileup.

\section{Conclusions}

There starts to be a certain choice of infrared and collinear safe
inclusive jet algorithms for hadron colliders. 
As we have seen, some of these ($k_t$ and
Cambridge/Aachen) belong to a more general class of sequential
recombination algorithms, parametrised by a continuous parameter $p$,
which sets the power of the transverse momentum scale relative to the
geometrical distance ($p=1$ gives $k_t$, $p=0$ gives
Cambridge/Aachen).

Rather surprisingly, taking $p$ to be negative also yields an
algorithm that is infrared and collinear safe and has sensible
phenomenological behaviour.
We have specifically studied $p=-1$, the ``$\akt$'' algorithm, and
highlighted various simple theoretical properties, notably the
resilience of its jet boundaries with respect to soft radiation. The
other properties that we've discussed are essentially consequences of
this feature. These properties are characteristic also of certain
``iterative cone'' (and fixed-cone) algorithms, those with progressive removal (IC-PR)
of the stable cones, as used for example in CMS.  However in the
$\akt$ algorithm these properties are 
obtained without having to pay the price of collinear
unsafety. Therefore the $\akt$ algorithm should be a good candidate as
a replacement algorithm for IC-PR algorithms.

One might worry that the resilience (or rigidity) with respect to soft
radiation could worsen the practical performance of the tool. This
seems not to be the case, at the least in examples examined so far,
including the one shown in section~\ref{sec:top}, 
top reconstruction. One reason for
this might be that any loss of resolution due to inflexibility in
adapting to perturbative branching may be compensated for by the
reduced fluctuations caused by the underlying event radiation (due to
suppression of area fluctuations and back reaction).
It should also be kept in mind that soft particles contribute only a
modest component of the overall jet momentum, and the algorithm remains
flexible in its adaptation to hard (sub)structure in the jets.
In this respect it might also be interesting in future
phenomenological investigations to examine less negative values of $p$, 
for which the `hard'-adaptability extends down to softer momenta than
is the case with $p=-1$.

\paragraph{Note.} Subsequent to the first presentation of the
anti-$k_t$ algorithm and its main properties at the June 2007 Les
Houches workshop on Physics at TeV colliders,
it was brought to our attention by Pierre-Antoine Delsart and Peter
Loch that a variant of the $k_t$ sequential recombination algorithm
had been coded within the ATLAS software framework and called
reverse-$k_t$. It has distance measures $d_{ij}=\max(k_{ti}^2,
k_{tj}^2) R^2/\Delta_{ij}^2$ and $d_{iB} =k_{ti}^2$, and recombines
the pair of objects with the largest $d_{ij}$ unless a $d_{iB}$ is
larger in which case $i$ is called a jet. We observe that these
distance measures are just the reciprocals of those for anti-$k_t$;
together with the recombination of the most distant pair first, this
causes the algorithm to produce identical jets to anti-$k_t$.

\section*{Acknowledgements}

One of us (GPS) thanks Bruce Knuteson for a stimulating discussion on
the interest of parametrised clustering jet algorithms.
We are grateful to G\"unther Dissertori for providing us with the
details of the implementation of the CMS iterative cone algorithm.
Figure~\ref{fig:top} was generated making use of tools developed in
collaboration with Juan Rojo and we thank him for discussions on this
and related matters.
This work has been supported in part by grant ANR-05-JCJC-0046-01 from
the French Agence Nationale de la Recherche and under Contract
No. DE-AC02-98CH10886 with the U.S. Department of Energy.

\appendix 
\section*{Appendix: comment on the ``iterative cone''}

There are two broad classes of iterative cone algorithms: some find
stables cones (of radius $R$) by iterating from all seeds (and
sometimes midpoint seeds) and then run some split--merge procedure
to deal with overlapping stable cones.  These algorithms,
iterative cones with a split--merge step (IC-SM), are used at the Tevatron 
(for instance the MidPoint algorithm) and it
is our understanding that the ATLAS cone jet algorithm is of this type too.
A second class takes the hardest seed particle in the event, iterates
to find a stable cone, calls it a jet, removes all particles contained
in that jet from the event, and then repeats the procedure with the
remaining particles, over and over until there remain no seeds. This
iterative-cone with progressive removal of particles (IC-PR) is the
iterative cone in the CMS experiment.\footnote{It is often ascribed to
  UA1, ref.~\cite{UA1}, however the algorithm described there is not
  based on the iteration of cones.} %

IC-SM algorithms have been widely studied, are known to suffer from
infrared (IR) safety issues (see \eg \cite{Blazey,EHHLT,siscone}). The
IR safe equivalent is SISCone.
Though referred to as ``cones,'' they do not as a rule give jets with
an area of $\pi R^2$ \cite{css-area}.

IC-PR algorithms have received less attention, and they behave rather
differently from the IC-SM variety. For reference we therefore
document some aspects of them here. A main point to note is that they
suffer from a \emph{collinear} safety issue (in the limit of a fine
calorimeter).
This can be illustrated with the following configuration for an
algorithm with radius $R$:
\begin{subequations}
\begin{align}
  p_1:\, y&=0,\, \phi =  0.0R,\, p_t = 130 \GeV\,,\\
  p_2:\, y&=0,\, \phi =  0.7R,\, p_t = 200 \GeV\,,\\
  p_3:\, y&=0,\, \phi =  1.5R,\, p_t = 90 \GeV\,.
\end{align}
\end{subequations}
The hardest seed is $p_2$, and in the small $R$ limit (in which the
results are independent of the recombination scheme) it leads to a jet
that contains all particles, and is centred at $\phi\simeq 0.65 R$.
If particle $p_2$ splits collinearly into 
\begin{subequations}
  \begin{align}
    p_{2a}:\, y&=0,\, \phi =  0.7R,\, p_t = 120 \GeV\,,\\
    p_{2b}:\, y&=0,\, \phi = 0.7R,\, p_t = 80 \GeV\,,
  \end{align}
\end{subequations}
then $p_1$ is now the hardest seed and it leads to a jet at $\phi \simeq
0.42R$ which contains $p_1$, $p_{2a}$, $p_{2b}$. That leaves $p_3$,
which seeds a second jet, and so the number of hard jets is modified
by the collinear splitting. This will for example lead to divergences
at NNNLO in inclusive jet cross sections, at NNLO in $3$-jet and
W+2-jet cross sections and at NLO in jet-mass distributions for
$3$-jet and W+2-jet events. We recall, as discussed \eg in
\cite{siscone}, that a cross section that diverges at N$^{p}$LO can
only be reliably calculated up to N$^{p-2}$LO. In this respect IC-PR
are `dangerous' to the same extent as midpoint-variants of the IC-SM
algorithms.\footnote{Note, however, that beyond LO in the inclusive
  jet spectrum, their perturbative expansions differ.}

Unlike their IC-SM cousins, and the collinear-safety issue aside,
IC-PR algorithms do have all the features of an ``ideal'' cone
algorithm as described here for the anti-$k_t$ algorithm, in particular they are
soft-resilient and give circular jets of radius $R$. Thus the
anti-$k_t$ algorithm seems a natural replacement for them, especially
as they happen to be identical perturbatively at NLO in the inclusive
jet spectrum (if used with the same recombination scheme).
An alternative replacement would be the following: find all stable
cones (as in SISCone), label the hardest one a jet, remove its
particles from the event, and then repeat the procedure on the
remaining particles, over and over. It too would have the property of
soft-resilience, but would differ perturbatively from IC-PR at
NLO. This, together with the issue of computational speed, leads us to
prefer the anti-$k_t$ algorithm as a replacement.


\begin{thebibliography}{99}


\bibitem{kt}
  S.~Catani, Y.~L.~Dokshitzer, M.~H.~Seymour and B.~R.~Webber,
  Nucl.\ Phys.\ B {\bf 406}  (1993)  187 and refs.\ therein;
  S.~D.~Ellis and D.~E.~Soper,
  Phys.\ Rev.\ D {\bf 48} (1993) 3160 
  [hep-ph/9305266]. 

\bibitem{cam}
  Y.~L.~Dokshitzer, G.~D.~Leder, S.~Moretti and B.~R.~Webber,
  JHEP {\bf 9708}, 001 (1997)
  [hep-ph/9707323];
  M.~Wobisch and T.~Wengler,
  hep-ph/9907280.

\bibitem{Blazey}
  G.~C.~Blazey {\it et al.},
  hep-ex/0005012.

\bibitem{css-area}
  M.~Cacciari, G.~P.~Salam and G.~Soyez,
      JHEP {\bf 0804} (2008) 005
        [arXiv:0802.1188].


\bibitem{siscone}
  G.~P.~Salam and G.~Soyez,
  JHEP {\bf 0705} (2007) 086 
  [arXiv:0704.0292].

\bibitem{CMSTDR}
  G.~L.~Bayatian {\it et al.}  [CMS Collaboration],
  ``CMS physics: Technical design report.''

\bibitem{Pythia}
  T.~Sjostrand {\it et al.},
  Comput.\ Phys.\ Commun.\  {\bf 135} (2001) 238 
  [hep-ph/0010017];
  T.~Sjostrand {\it et al.},
  hep-ph/0308153.

\bibitem{Herwig}
  G.~Corcella {\it et al.},
  arXiv:hep-ph/0210213.

\bibitem{DasSalJetMass}
  M.~Dasgupta and G.~P.~Salam,
  Phys.\ Lett.\  B {\bf 512}, 323 (2001)
  [arXiv:hep-ph/0104277].

\bibitem{DasSalEFlow}
  M.~Dasgupta and G.~P.~Salam,
  JHEP {\bf 0203} (2002) 017
  [arXiv:hep-ph/0203009].

\bibitem{BKS}
  C.~F.~Berger, T.~Kucs and G.~Sterman,
  Phys.\ Rev.\  D {\bf 68}  (2003) 014012
  [arXiv:hep-ph/0303051].

\bibitem{BMS}
  A.~Banfi, G.~Marchesini and G.~Smye,
  JHEP {\bf 0208}  (2002) 006
  [arXiv:hep-ph/0206076].

\bibitem{ApplebySeymour}
  R.~B.~Appleby and M.~H.~Seymour,
  JHEP {\bf 0212} (2002) 063 
  [arXiv:hep-ph/0211426].

\bibitem{BanfiDasgupta}
  A.~Banfi and M.~Dasgupta,
  JHEP {\bf 0401} (2004) 027 
  [arXiv:hep-ph/0312108];
  Y.~Delenda, R.~Appleby, M.~Dasgupta and A.~Banfi,
  JHEP {\bf 0612} (2006) 044 
  [arXiv:hep-ph/0610242].

\bibitem{DLMS}
  Y.~L.~Dokshitzer, A.~Lucenti, G.~Marchesini and G.~P.~Salam,
  Nucl.\ Phys.\  B {\bf 511}, 396 (1998)
  [Erratum-ibid.\  B {\bf 593}, 729 (2001)]
  [arXiv:hep-ph/9707532];
  JHEP {\bf 9805}, 003 (1998)
  [arXiv:hep-ph/9802381];
  M.~Dasgupta and B.~R.~Webber,
  JHEP {\bf 9810} (1998) 001
  [arXiv:hep-ph/9809247].

\bibitem{DasMagSmy}
  M.~Dasgupta, L.~Magnea and G.~Smye,
  JHEP {\bf 9911} (1999) 025
  [arXiv:hep-ph/9911316].

\bibitem{BenBroMag}
  M.~Beneke, V.~M.~Braun and L.~Magnea,
  Nucl.\ Phys.\  B {\bf 497} (1997) 297
  [arXiv:hep-ph/9701309].

\bibitem{LeeSterman}
  C.~Lee and G.~Sterman,
  Phys.\ Rev.\  D {\bf 75}  (2007) 014022
  [arXiv:hep-ph/0611061].

\bibitem{fastjet}
  M.~Cacciari and G.~P.~Salam,
  Phys.\ Lett.\ B {\bf 641} (2006) 57
  [arXiv:hep-ph/0512210].


\bibitem{CGALTriang}
A.~Fabri {\it et al.},
Softw.~Pract.~Exper.~ {\bf 30} (2000) 1167;
J.-D.~Boissonnat {\it et al.},
Comp.~Geom.~{\bf 22} (2001) 5.


\bibitem{FastJetWeb} M.~Cacciari, G.~P.~Salam and G.~Soyez,
  \url{http://www.lpthe.jussieu.fr/~salam/fastjet}

\bibitem{UA1}
  G.~Arnison {\it et al.}  [UA1 Collaboration],
  Phys.\ Lett.\  B {\bf 132} (1983) 214.

\bibitem{EHHLT}
  S.~D.~Ellis, J.~Huston, K.~Hatakeyama, P.~Loch and M.~Tonnesmann,
      Prog.\ Part.\ Nucl.\ Phys.\  {\bf 60} (2008) 484
        [arXiv:0712.2447].

\end{thebibliography}
\end{document}